\documentstyle[12pt]{article}
\newcommand{\PSbox}[3]{\mbox{\rule{0in}{#3}\includegraphics{#1}\hspace{#2}}}
\newcommand{\beq}{\begin{eqnarray}}
\newcommand{\eeq}{\end{eqnarray}}
\makeatletter
\def\vereq#1#2{\lower3pt\vbox{\baselineskip1.5pt \lineskip1.5pt
\ialign{$\m@th#1\hfill##\hfil$\crcr#2\crcr\sim\crcr}}}
\makeatother

\begin{document}

\begin{titlepage}
\begin{center}
    \hfill    LBNL-44151 \\
    \hfill    UCB-PTH-99/34 \\
    \hfill    PUPT-1885  \\
    \hfill    NSF-ITP-99-100\\
{}~{} \hfill hep-th/9908186\\
\vskip .3in
{\Large \bf Brane Junctions in the Randall-Sundrum Scenario}

\vskip 0.3in
{\bf Csaba Cs\'aki$^{a,b,}$\footnote{Address after September 1:
Theoretical Division T-8, Los Alamos National Laboratory, Los Alamos,
NM 87545.} and Yuri Shirman$^{c,d}$}

\vskip 0.15in

$^a${\em Department of Physics\\
University of California, Berkeley, CA 94720}

\vskip 0.1in

$^b${\em Theoretical Physics Group\\
     Ernest Orlando Lawrence Berkeley National Laboratory\\
     University of California, Berkeley, CA 94720}

\vskip 0.1in

$^c${\em Department of Physics\\
Princeton University, Princeton, NJ 08544}

\vskip 0.1in

$^d${\em Institute for Theoretical Physics\\
University of California, Santa Barbara, CA 93106}

\vskip 0.1in
{\tt  ccsaki@lbl.gov, yuri@feynman.princeton.edu}

\end{center}

\vskip .25in
\begin{abstract} 
We present static solutions to Einstein's equations corresponding to
branes at various angles intersecting in a single 3-brane. Such
configurations may be useful for building models with localized
gravity via the Randall-Sundrum mechanism. We find that such solutions
may exist only if the mechanical forces acting on the junction exactly
cancel. In addition to this constraint there are further conditions
that the parameters of the theory have to satisfy. We find that at
least one of these involves only the brane tensions and cosmological
constants, and thus can not have a dynamical origin. We present these
conditions in detail for two simple examples.

We discuss the nature of the cosmological constant problem in the
framework of these scenarios, and outline the desired features of the
brane configurations which may bring us closer towards the resolution
of the cosmological constant problem. 
\end{abstract}
\end{titlepage}

\newpage
\section{Introduction}
\setcounter{equation}{0}
\setcounter{footnote}{0}
The principal challenge facing particle theorists is to understand the 
physics at energy scales of a few TeV. It seems inevitable that the
standard model be amended at these scales. The most popular scenario
is that the world is supersymmetric, with the scale of supersymmetry
breaking being around a few hundred GeV. Thus in this scenario all 
superpartners would become visible around the TeV scale. This
possibility would explain why there is such a big hierarchy between
the weak and the Planck scales. Thus the bulk of the efforts in the
past twenty years has been devoted to modifying particle physics above 
the weak scale in order to accommodate this huge hierarchy. Very
recently it has been understood that there exists a different way
towards reconciling particle physics with gravity at high energies, by 
radically changing our ideas how gravity will work above the TeV
scale~\cite{Nima,otherextra,HW}. 
Most notably, Arkani-Hamed, Dimopoulos and Dvali suggested~\cite{Nima},
that in fact the fundamental Planck scale itself could be as low as a
few TeV, if there are large extra dimensions. This way the 
problem of the hierarchy between the Planck and the weak scales is
translated into the question of why the size of the extra dimensions
is much larger than its natural scale of 1/TeV. The fundamental new
ingredient in this idea is that the reason why we do not see the
effects of the large extra dimensions is because the standard model
fields live on a 3-brane, and the only fields which can propagate in
the extra dimensions are the gravitons. 

Recently, Randall and Sundrum (RS) further developed on these ideas by
noting, that our understanding of Kaluza-Klein (KK) gravity models has 
been largely limited to factorizable metrics where the components of
the metric tensor do not depend on the coordinates of the extra
dimension~\cite{randall,randall2}. 
RS noted that if this is not the case, the properties of
compactification may change radically. In particular~\cite{randall}, 
following the
idea that the standard model fields may live on a 3-brane, RS
considered two 3-branes embedded into 4+1 dimensional spacetime, with
the extra dimension being a compact $S^1/Z_2$ manifold (this latter
motivated by \cite{HW}). The bulk cosmological constant was chosen to
be negative, while the tensions of the two branes are of opposite
signs. RS found that if a particular fine-tuning relation between the
cosmological constant and the brane tensions is obeyed, there will be
a static solution to Einstein's equations, which is given by two
slices of Anti-de Sitter (AdS) space glued together at the location of 
the branes. The metric tensor has a non-trivial exponential dependence 
on the coordinate $y$ along the extra dimension\footnote{Similar domain-wall 
solutions in the context of supergravity theories have been considered 
in~\cite{Cvetic}.}. 
This exponential
determines the natural mass scale at the location $y$. Thus it is not
inconceivable, that while the mass scale at the brane with positive
tension is $10^{19}$ GeV, due to the exponential suppression it might
be a few TeV on the brane with negative tension, thereby possibly
solving the hierarchy problem~\cite{randall,randall-lykken}. 
RS further noted~\cite{randall2}, that the brane with
positive tension supports a single bound state (zero mode) of
gravitons, thereby ``trapping'' gravity to this wall. This is a very
appealing feature of the theory, since in this case one might as well
move the second brane with negative tension far away (in fact
making the size of the extra dimension infinitely large), while
Newton's law of gravity is still correctly reproduced on the brane due 
to the trapped zero mode. The idea of having non-compact extra
dimensions is also explained in Refs.~\cite{noncompact,Gogberashvili}
Since the trapping of the zero mode
crucially depends on the fact that one has a brane of co-dimension one, one
would think that this feature of trapping gravity on a 3-brane can
only hold if one has a 4+1 dimensional spacetime. However,
Arkani-Hamed, Dimopoulos, Dvali and Kaloper have pointed
out~\cite{ADDK}, that if
one considers intersecting branes of co-dimension one (intersecting
orthogonally in a single 3-brane) one can still find static solutions
to Einstein's equations, which will trap gravity to the intersection
of the branes. Further solutions to Einstein's equation have been
given in \cite{Oda}, within the context of supergravity in 
~\cite{Kehagias,Kostas}, and the relation to string theory and
holography has been explained in~\cite{verlinde}. The cosmological
aspects of the RS models have been studied
in~\cite{cosmology,kaloper}, while the issue of bulk scalars and
stabilization of the radius in~\cite{Wise1,Wise2}.

In  this paper we consider more general intersections of branes. In
particular, we discuss ``brane junctions'', that is intersections of
semi-infinite branes intersecting in a single 3-brane. We will mainly
concentrate on junctions of 4-branes, but we expect that it will be
straightforward to generalize the algorithm of gluing sectors of
static AdS spacetimes together to higher dimensions. We find, that
brane junctions can yield static solutions to Einstein's equations 
if some fine-tuning conditions between the tensions and the
cosmological constants are satisfied. Moreover, the balance of
mechanical forces on the junction arising from the brane tensions
is a necessary condition for the existence of the static
solution.
We present these conditions for some simple examples in
detail.

A crucial ingredient of the RS solution is the fine-tuning between the 
brane tension and the bulk cosmological constant, which insures that
there is a static universe with the effective 4-dimensional
cosmological constant vanishing. Thus the cosmological constant
problem in four dimensions is translated into the problem of the
tuning between the brane tension and the fundamental (five
dimensional) cosmological constant. In the case of branes intersecting 
at angles one expects that there will be similar relations, which also 
involve the angles of the branes. A simple way of understanding the
cosmological constant problem would then be to imagine that one starts 
with a setup of branes whose angles do not satisfy the required tuning 
relation. Then one lets the system relax, and perhaps it would
settle to a configuration where the angles of the branes take the
right value, thus providing flat 4 dimensional universe with a vanishing 
cosmological constant. For this scenario to be viable, one would need
to find a solution of intersecting branes, where all fine-tuning
conditions can be satisfied by the choice of angles between the
branes. Moreover, this configuration should be a ground state of the
system once the dynamics of the branes is included.
Unfortunately, as we will see, this is not the case in the
solutions based on junctions presented in this paper. There is always
at least one remaining fine-tuning involving only the tensions and the 
cosmological constants. One may hope however, that a more clever
configuration of branes may posses the necessary features and thus provide 
a dynamical interpretation of the cosmological constant problem. 

This paper is organized as follows: in Section 2 we review the RS
solution by considering a 3-brane in 4+1 dimensional spacetime
separating two domains with different cosmological constants. In
Section 3 we give our general setup for brane junctions in 5+1
dimensions and discuss the general algorithm of finding the solutions to
Einstein's equations and the fine-tuning relations. In Section 4 we
work out the solutions and fine-tuning relations in detail for two
simple junctions. In Section 5 we summarize our observations about the 
cosmological constant problem, and we conclude in Section 6.

\section{Review of the Randall-Sundrum Solution}
\setcounter{equation}{0}
\setcounter{footnote}{0}

\begin{figure}
\PSbox{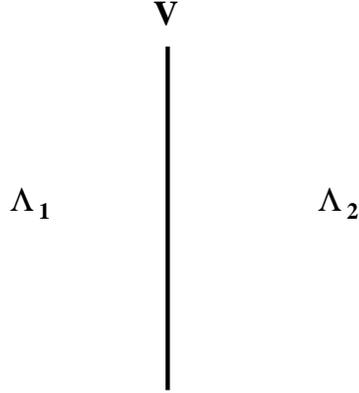 hscale=60 vscale=60 hoffset=150  voffset=0}{8cm}{5cm}
\caption{A single 3-brane with tension $V$ divides the 4+1
dimensional space-time into two domains with different cosmological constants.
\label{fig:domains}}
\end{figure}

We first briefly review the original Randall-Sundrum (RS) solution by
presenting a slightly generalized version of it. In this setup we have
a single 3-brane (with positive tension $V$) embedded into  4+1
dimensional spacetime, where the branes divide the space into two
domains: one with cosmological constant $\Lambda_1$, the other with
$\Lambda_2$ (both of them negative). This setup is depicted in 
Fig.~\ref{fig:domains}.
The original RS solution for
$\Lambda_1=\Lambda_2=\Lambda$ is given by
\begin{equation}
ds^2=e^{-2m|y|}\eta_{ab}dx^adx^b-dy^2,
\label{eq:RS}
\end{equation}
where $a,b=0,1,2,3$ are the coordinates of the four dimensional
spacetime, while $y$ is the coordinate along the (infinite) extra  
dimension. In order for this to be the solution to the Einstein
equations, the parameter $m$ has to satisfy
\begin{equation}
m^2=-\frac{\kappa^2\Lambda}{6},
\label{eq:bulk}
\end{equation}
where $\kappa^2$ is Newton's constant in five dimensions
($\kappa^2=\frac{1}{M_*^3}$ where $M_*$ is the five dimensional Planck 
scale),
and the tension of the brane has to be tuned to be
\begin{equation}
V=\sqrt{-\frac{6\Lambda}{\kappa^2}}.
\label{eq:jump}
\end{equation}
For the generalizations to be presented below it turns out to be
useful following \cite{ADDK}
to redefine the coordinates such that one obtains a conformally
flat metric:
\begin{equation}
dy=e^{-m|y|}dz
\label{eq:redefine}
\end{equation}
In these coordinates 
\begin{equation}
ds^2=\omega^2 (z) \eta_{\mu\nu} dx^{\mu}dx^{\nu},
\label{eq:conformal}
\end{equation}
where
\begin{equation}
\omega^{-1}(z)=m|z|+1,
\label{eq:omega}
\end{equation}
if one wants to have the location of the brane to be at $z=0$. In
these coordinates it is easy to see why (\ref{eq:conformal}) solves
the Einstein equations with a negative cosmological constant $\Lambda$
and a brane with tension $V$ at $z=0$. 

The Einstein tensor for a metric of the form
$g_{\mu\nu}=\omega^2 \tilde{g}_{\mu\nu}$ in $d$ dimensions 
is given by 
\begin{equation}
G_{\mu\nu}=\tilde{G}_{\mu\nu}+
(d-2)\left[ \tilde{\nabla}_{\mu}\log\omega\tilde{\nabla}_{\mu}\log\omega-
\tilde{\nabla}_{\mu}\tilde{\nabla}_{\nu}\log\omega+\tilde{g}_{\mu\nu}
\left( 
\tilde{\nabla}^2
\log\omega+\frac{d-3}{2} (\tilde{\nabla}\log\omega
)^2 
\right) 
\right].
\label{eq:Einstein}
\end{equation}
where the covariant derivatives $\tilde{\nabla}$ are evaluated with
respect to the metric $\tilde{g}$. Since in our case the metric is
conformally flat, $\tilde{g}_{\mu\nu}=\eta_{\mu\nu}$, all covariant
derivatives can be replaced by ordinary derivatives, and for the same reason
$\tilde{G}_{\mu\nu}=0$. For the case $\omega^{-1}(z)=m|z|+1$
one can easily see that the Einstein equations at an arbitrary point
of the bulk ($z\neq 0$) are satisfied if $6m^2=-\kappa^2\Lambda$, since
the energy-momentum tensor in the bulk is given by
$T_{\mu\nu}^{bulk}=\Lambda\eta_{\mu\nu} \omega^2(z)$. The
singularities in the second derivatives of $\omega$ result in the
additional term
\begin{equation}
6 m\omega (z) \delta (z)\,
{\rm diag} (1,-1,-1,-1,0)
\label{eq:sing}
\end{equation}
in the Einstein tensor, which must be balanced by the term from 
energy-momentum tensor of the brane on the right hand side of
Einstein's equations 
\begin{equation}
\kappa^2\omega (z) V\delta (z)\, 
{\rm diag} (1,-1,-1,-1,0),
\end{equation}
thus yielding $6m=\kappa^2 V$.

This solution represents two slices of Anti-de Sitter space (the
solution of Einstein's equations with negative cosmological constant)
glued together at $z=0$. The brane represents the source necessary for
fitting the two pieces together. Now it is trivial to generalize
this solution to the case with two domains with different cosmological
constants. It is a space with two slices of AdS spaces with different
$m$'s glued together. Thus one expects that a conformally flat metric
(\ref{eq:conformal}) with 
\begin{equation}
\omega^{-1}(z)=m_1z\theta (z)-m_2z\theta (-z)+1,
\label{eq:domains}
\end{equation}
where $\theta (z)=1$ for $z>0$ and $\theta (z)=0$ for $z<0$ is the 
Heaviside step-function. Einstein's equations in the bulk require that
\begin{equation}
m_1^2=-\frac{\kappa^2\Lambda_1}{6},\; \; \;
m_2^2=-\frac{\kappa^2\Lambda_2}{6}, 
\label{eq:bulk2}
\end{equation}
and the tension of the brane is determined by
\begin{equation}
3(m_1+m_2)=\kappa^2V.
\label{eq:jump2}
\end{equation}
Thus the fine-tuning condition in this case is given by
\begin{equation}
\kappa^2V^2=\frac{3}{2}(\sqrt{-\Lambda_1}+\sqrt{-\Lambda_2})^2.
\label{eq:tune}
\end{equation}

Clearly by construction the solution we found is static. However,
we included the brane as an internal source nailed at $z=0$. The
dynamics of the brane is not included in this simple description,
and thus it is impossible to determine if the solution is stable
against small fluctuations.

The above example already suggests how one can further generalize
these solutions by fitting slices of AdS space with different
cosmological constants together. Indeed, Arkani-Hamed, Dimopoulos,
Dvali and Kaloper have showed, that one can find solutions
corresponding to orthogonally intersecting branes. In the next section
we show that one can also find solutions corresponding to the junction
of semi-infinite branes intersecting in a single 3-brane. We will
concentrate on the case of 4-branes embedded in 5+1 dimensional
spacetime, but we expect that generalizations to higher dimensions
based on the algorithm described below should be straightforward.

\section{The General Setup}
\setcounter{equation}{0}
\setcounter{footnote}{0}
\begin{figure}
\PSbox{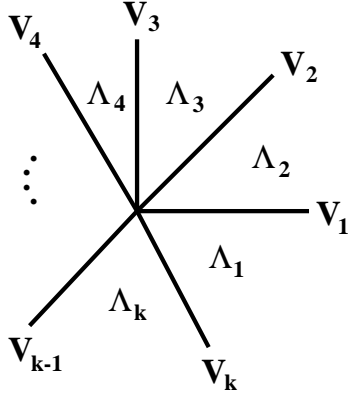 hscale=60 vscale=60 hoffset=150  voffset=0}{8cm}{5cm}
\caption{The setup of semi-infinite 4-branes intersecting in a
single 3-brane in 5+1 dimensions. The brane tensions are denoted by
$V_i$, while the bulk cosmological constants are given by $\Lambda_i$. 
\label{fig:general}}
\end{figure}

We consider a junction of half (semi-infinite in one direction)
4-branes in 5+1 spacetime dimensions. These branes intersect in a
single 3-brane, and the tension of the i$^{th}$ brane is $V_i$. The 
bulk cosmological constant in the region between the i$^{th}$ and
(i+1)$^{st}$ brane is taken to be $\Lambda_i$. This general setup is
depicted in Fig.~\ref{fig:general}. We want to fit slices of static 
5+1 dimensional Anti-de Sitter (AdS) space together such that the resulting
full solution exactly corresponds to the setup given in
Fig.~\ref{fig:general}. A patch of 5+1 dimensional AdS space can be
described by the conformally flat metric 
\begin{equation}
ds^2= \omega^2(x,y) \eta_{\mu \nu}dx^{\mu}dx^{\nu},
\label{eq:AdSgeneral}
\end{equation}
where $x_{0,1,2,3}$ are the coordinates of the 4 dimensional spacetime, and
$x_4\equiv x$, $x_5\equiv y$ are the coordinates in the extra
dimensions. The conformal factor is given by
\begin{equation}
\omega^{-1}(x,y)= \vec{m} \cdot \vec{x}+1,
\end{equation}
where $\vec{x}=(x,y)$, the parameters
$\vec{m}=(m_x, m_y)$ are related to (negative) the bulk cosmological
constant $\Lambda$ as $m_x^2+m_y^2=-\frac{\kappa^2}{10} \Lambda$, and
$\kappa^2$ is Newton's constant in six dimensions
($\kappa^2=1/M_{*}^4$, where $M_{*}$ is the fundamental Planck scale
in six dimensions). Note that the requirement that the conformal
factor $\omega$ is positive imposes certain inequalities on the 
possible values of $\vec{m}$ in each AdS patch.

In order to find the full solution to Einstein's equations we need to
glue the $\omega$'s together such that 

- the metric tensor is continuous at the location of the branes

- the discontinuity in the derivatives along the branes reproduces the 
energy momentum tensor of the brane with given tension $V_i$ rotated
into the appropriate direction.

It is convenient to write the conformal factor in a space
composed of $k$ AdS patches as
\beq
\label{eq:omegatotal}
\omega^{-1}= 1+ \sum_{i=1}^k  (\vec{m}_i \cdot \vec{x})\,
\theta (\vec{n}_{i-1} \cdot \vec{x})\,
\theta (-\vec{n}_{i} \cdot \vec{x})  \ ,
\eeq
where $\vec{n}_i=(- \sin \varphi_i, \cos \varphi_i)$ is a unit
vector in the $x_4-x_5$ plane
normal to the i$^{th}$ brane, and $\varphi_i$ is the angle
between the brane and the coordinate axis. Clearly, one linear
combination of angles is an unphysical parameter corresponding to
the overall rotation of the configuration. Thus we can choose the
coordinate system such that $\varphi_1=0$. We conclude that the
ansatz (\ref{eq:omegatotal}) depends on $k$ vectors $\vec{m}_i$
and $k-1$ angles, altogether $3k-1$ parameters.

We now turn to the energy-momentum tensor of the configuration of
$k$ AdS patches separated by branes. In the bulk of a
given patch the energy momentum tensor is given by
$T_{\mu \nu}^{bulk,i} = \Lambda_i \omega^2 \eta_{\mu\nu}$. Thus
at the generic point the energy-momentum tensor can be written as
\beq
T_{\mu \nu}^{bulk} =
\sum_{i=1}^k\Lambda_i\, \omega^2\, \theta (\vec{n}_{i-1} \cdot \vec{x})
\, \theta (-\vec{n}_{i} \cdot \vec{x})\,\eta_{\mu\nu} \ .
\eeq

The energy-momentum tensor of a 4-brane rotated by an angle $\varphi$
from the horizontal direction $x$ is given by
\begin{equation}
T_{\mu \nu}^{brane,i}= V_i\, \omega(x,y) \,
\delta (\vec{n}_i \cdot \vec{x}) 
\left( \begin{array}{cccccc}
1\\&-1\\&&-1\\&&&-1\\&&&& -\cos^2 \varphi_i & -\sin \varphi_i \cos \varphi_i \\
&&&& -\sin \varphi_i \cos \varphi_i & -\sin^2 \varphi_i \end{array} \right ).
\label{eq:Tmunu}
\end{equation}
Thus the total stress-energy tensor in our space is given by
\beq
\label{eq:Tmunutotal}
T_{\mu\nu} = T_{\mu\nu}^{bulk} + \sum_{i=1}^k
T_{\mu\nu}^{brane,i}
\eeq

The Einstein tensor $G_{\mu\nu}=R_{\mu\nu}-\frac{1}{2}g_{\mu\nu}R$ for
a conformally flat metric $g_{\mu\nu}=\omega^2 \eta_{\mu\nu}$ in $d$
dimensions is given by
\begin{equation}
G_{\mu\nu}=(d-2)\left[ \partial_{\mu}\log\omega\partial_{\mu}\log\omega-
\partial_{\mu}\partial_{\nu}\log\omega+\eta_{\mu\nu} \left( \partial^2
\log\omega+\frac{d-3}{2} (\partial \log\omega )^2 \right) 
\right].
\label{eq:Einstein2}
\end{equation}

We are now ready to solve the Einstein equations. At a generic
point in the bulk we find
\begin{equation}
\vec{m}_i^2=-\frac{\kappa^2}{10} \Lambda_i.
\label{eq:bulk3}
\end{equation}
The requirements that the singularities in the derivatives at the
brane reproduce the brane tension will yield two equations at
each brane\footnote{This is easy to see by going to a coordinate
system in which the brane under consideration is horizontal, so
that the relevant parts of both the energy-momentum and the Einstein
tensors are diagonal.}:
\beq
\Delta \vec{m}_i = \vec{m}_{i+1} - \vec{m}_i =
\frac{\kappa^2 V_i}{4} \vec{n}_i \ .
\label{eq:boundary}
\eeq
To summarize, we found $3k$ equations on the
$3k-1$ parameters of the ansatz
(\ref{eq:AdSgeneral}), (\ref{eq:omegatotal}).
Therefore, generically the 
$k$ bulk cosmological constants $\Lambda_i$ and the $k$
brane tensions $V_i$ need to satisfy a single (but quite
complicated) fine-tuning condition. We will discuss this fine
tuning condition in more detail in the particular
examples  in the
following section.
Once this fine-tuning
condition is satisfied a static solution of the form
(\ref{eq:AdSgeneral}), (\ref{eq:omegatotal}) exists and its
parameters are completely determined.

It is worth noting that (as
should have been expected) the solution satisfies the requirement
that (classical) mechanical forces 
acting at the junction exactly balance. Indeed summing up
equations (\ref{eq:boundary}) we find
\beq
\sum_{i=1}^k V_i \, \vec{n}_i=0 \ ,
\eeq
which can be rewritten as
\beq
\sum_{i=1}^k \vec{V}_i = 0 \ ,
\eeq
where $\vec{V}_i = (V_{x,i}, V_{y,i}) = (V_i \cos \varphi_i, V_i
\sin \varphi_i)$. The latter equation is exactly the condition of
vanishing force.

\section{Examples}
\setcounter{equation}{0}
\setcounter{footnote}{0}
Below we will apply the formalism presented in the previous section to 
discuss two particular examples in detail. The first example will
involve two 4-branes intersecting at an angle, with different bulk
cosmological constants in the four domains of spacetime, while the
second example will involve three semi-infinite 4-branes intersecting 
in a single 3-brane (a ``triple junction''). We will give the
necessary fine-tuning conditions in detail, and find the metric tensor 
in every sector of spacetime.

\subsection{4-branes intersecting at an angle}

\begin{figure}
\PSbox{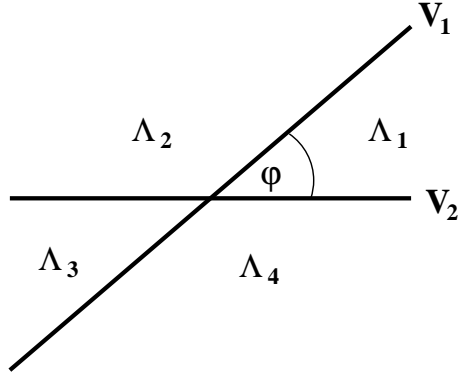 hscale=60 vscale=60 hoffset=150  voffset=0}{8cm}{5cm}
\caption{Two 4-branes with tensions $V_1$ and $V_2$
intersecting at an angle $\varphi$. The four domains may have different
cosmological constants.
\label{fig:angle}}
\end{figure}

In our first example we will consider two 4-branes embedded into a 5+1
dimensional spacetime. The tensions of the two branes are given by
$V_1$ and $V_2$, and the four domains may have different cosmological
constants. The setup is given in Fig.~\ref{fig:angle}. Note, that
since we are considering infinite 4-branes the condition on the forces 
balancing at the junction is automatically satisfied, thus at this
point the angle $\varphi$ between the branes is arbitrary.

Following the general formalism of the previous section, we write the
metric in the form $g_{\mu\nu}=\omega^2 (x,y) \eta_{\mu\nu}$, where
\begin{eqnarray}
\omega^{-1}(x,y)= && f_1(x,y) \theta (y)\theta (x\cos \varphi-y\sin\varphi) +
f_2(x,y) \theta (y)\theta (y\sin\varphi-x\cos \varphi)
+\nonumber \\
&& f_3(x,y) \theta (-y )\theta (y\sin\varphi-x\cos \varphi) +
f_4(x,y) \theta (-y)\theta (x\cos \varphi-y\sin\varphi)  +1, \nonumber \\
\label{eq:fourdomains}
\end{eqnarray}
where $1,2,3,4$ label the four domains where the value of the
cosmological constant is $\Lambda_{1,2,3,4}$, and the $f_{i}(x,y)$ are 
functions linear in $x,y$ and positive everywhere inside the domain
\begin{equation}
f_{i}(x,y)=m_{i,x} x+m_{i,y} y.
\label{eq:fs}
\end{equation}
The Einstein equations in the bulk result in the conditions
\begin{eqnarray}
&m_{1x}^2+m_{1y}^2=-\lambda_1,\; \; \; 
 &m_{2x}^2+m_{2y}^2=-\lambda_2, \nonumber \\
&m_{3x}^2+m_{3y}^2=-\lambda_3, \; \; \;  
&m_{4x}^2+m_{4y}^2=-\lambda_4.
\end{eqnarray}
where we have used the notation $\lambda_i=\frac{\kappa^2}{10} \Lambda_i$.
The Einstein equations at the position of the branes will give the
conditions
\begin{eqnarray}
& m_{2y}-m_{1y}=v_1 \cos\varphi ,  \; \; \;
& m_{1x}-m_{2x}=v_1 \sin\varphi , \nonumber \\
& \hspace*{-1cm}m_{2y}-m_{3y}=v_2, \; \; \;
& m_{3x}-m_{2x}=0, \nonumber \\
& m_{3y}-m_{4y}=v_1 \cos\varphi , \; \; \;
& m_{4x}-m_{3x}=v_1 \sin \varphi , \nonumber \\
& \hspace*{-1cm} m_{1y}-m_{4y}=v_2, \; \; \;
& m_{4x}-m_{1x}=0,
\end{eqnarray}
where we have used the notation $v_i=\frac{\kappa^2}{4} V_i$.
We can express all variables with the help of $m_{1x}$, $m_{1y}$,
and $\varphi$
using the discontinuity equations as
\begin{eqnarray}
& m_{2x}=m_{1x}-v_1\sin\varphi ,\; \; \; & m_{2y}=m_{1y}+v_1\cos\varphi ,
\nonumber \\ 
& m_{3x}=m_{1x}-v_1\sin\varphi , \; \; \;&
m_{3y}=m_{1y}-v_2+v_1\cos\varphi ,
\nonumber \\
& \hspace*{-1.8cm} m_{4x}=m_{1x} ,\; \; \;& m_{4y}=m_{1y}-v_2.
\end{eqnarray}
Using these expressions the equations in the bulk can be rewritten as
\begin{eqnarray}
& \hspace*{-1cm} m_{1x}^2+m_{1y}^2=-\lambda_1, \; \; \; & (m_{1x}-v_1\sin\varphi )^2+
(m_{1y}+v_1\cos \varphi )^2=-\lambda_2, \nonumber \\
& m_{1x}^2+(m_{1y}-v_2)^2=-\lambda_4, & \; \; \;
\hspace*{-0.3cm} (m_{1x}-v_1\sin\varphi )^2+
(m_{1y}-v_2+v_1\cos \varphi )^2=-\lambda_3.
\label{eq:quadratic}
\end{eqnarray}
From the equations involving $\lambda_{1}$ and $\lambda_4$ we learn
that 
\begin{equation}
\label{eq:m1y}
m_{1y}=\frac{\lambda_4-\lambda_1+v_2^2}{2v_2}.
\end{equation}
Plugging this back into the other two equations and eliminating
$m_{1x}$ we get that
\begin{equation}
\cos \varphi= \frac{\lambda_3-\lambda_2+\lambda_1-\lambda_4}{2v_1v_2}=
\frac{2}{5}\frac{(\Lambda_3-\Lambda_2+\Lambda_1-\Lambda_4)}{\kappa^2 V_1V_2}.
\label{eq:phi}
\end{equation}
In particular, this relation implies, that in the case when the bulk
cosmological constant is isotropic
($\Lambda_1=\Lambda_2=\Lambda_3=\Lambda_4$) the only possible angle
between the branes is $\pi /2$.
The converse, however is not
true, and branes can be orthogonal with cosmological constants
different in each sector.
We now have two different expressions for $m_{1x}$ which can be
obtained from (\ref{eq:quadratic}).
Equating them
and substituting the values (\ref{eq:m1y}) for $m_{1y}$ and
(\ref{eq:phi}) for $\cos\varphi$ we obtain the fine-tuning condition
\begin{eqnarray}
&& (\lambda_1-\lambda_2+\lambda_3-\lambda_4)
(\lambda_1\lambda_3-\lambda_2\lambda_4) 
+v_2^2 (\lambda_1-\lambda_2)(\lambda_3-\lambda_4)
+v_1^2 (\lambda_1-\lambda_4)(\lambda_3-\lambda_2) \nonumber \\
&& -(\lambda_1+\lambda_2+\lambda_3+\lambda_4) v_1^2 v_2^2
-v_1^2 v_2^2 (v_1^2+v_2^2)=0.
\end{eqnarray}
Note, that the first three terms vanish if all cosmological constants
are set to be equal, and one is left with the fine-tuning
equation $-2 \lambda =v^2$, implying $\kappa^2 V^2=-\frac{16}{5}
\Lambda$, which exactly reproduces the fine-tuning condition obtained
in \cite{ADDK}.
Thus we find that the existence of the static solution determines
the angle between branes uniquely, and moreover, there is one
fine-tuning condition involving the cosmological constants and
the brane tensions.
For simplicity in our discussion we considered a specific case
of infinite branes. Have we considered semi-infinite branes with
different tensions, the solution would still exist subject to a
single (although more complicated) fine-tuning condition.

\subsection{Triple junction of semi-infinite 4-branes}

\begin{figure}
\PSbox{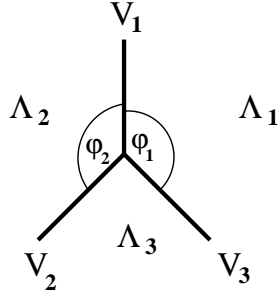 hscale=60 vscale=60 hoffset=175  voffset=0}{8cm}{5cm}
\caption{Three semi-infinite 4-branes intersecting at angles
  $\varphi_1$ and $\varphi_2$ in a single 3-brane.
\label{fig:triple}}
\end{figure}

In our second example we will consider three semi-infinite 
4-branes embedded into a 5+1
dimensional spacetime, intersecting in a single 3-brane. The setup is
depicted in Fig.~\ref{fig:triple}.
Similarly to the previous example, we write the metric in the form
 $g_{\mu\nu}=\omega^2 (x,y) \eta_{\mu\nu}$, where
\begin{eqnarray}
\omega^{-1}(x,y)= && f_1(x,y) \theta (x)\theta (y\sin\varphi_1-x\cos  
\varphi_1) + \nonumber \\
&& f_2(x,y) \theta (-x )\theta (y\sin\varphi_2+x\cos
\varphi_2)+\nonumber \\
&& f_3(x,y) \theta (x\cos \varphi_1-\sin\varphi_1)\theta
(-y\sin\varphi_2- x\cos\varphi_2), 
\label{eq:triple}
\end{eqnarray}
where $1,2,3$ label the three domains where the value of the
cosmological constant is $\Lambda_{1,2,3}$, and the $f_{i}(x,y)$ are 
functions linear in $x,y$ and positive everywhere inside the domain
\begin{equation}
f_{i}(x,y)=m_{i,x} x+m_{i,y} y.
\label{eq:fs2}
\end{equation}
The Einstein equations in the bulk are given by
\begin{eqnarray}
&&m_{1x}^2+m_{1y}^2=-\lambda_1, \nonumber \\
&&m_{2x}^2+m_{2y}^2=-\lambda_2, \nonumber \\
&&m_{3x}^2+m_{3y}^2=-\lambda_3,
\end{eqnarray}
where we have again 
used the notation $\lambda_i=\frac{\kappa^2}{10} \Lambda_i$.
The Einstein equations at the position of the branes will give the
conditions
\begin{eqnarray}
& \hspace*{-1.3cm} m_{2y}-m_{1y}=0, \; \; \; 
& m_{1x}-m_{2x}=v_1, \nonumber \\
& m_{2y}-m_{3y}=v_2 \sin\varphi_2, \; \; \;
& m_{2x}-m_{3x}=v_2 \cos\varphi_2, \nonumber \\
& m_{1y}-m_{3y}=v_3\sin\varphi_1, \; \; \;
& m_{3x}-m_{1x}=v_3 \cos\varphi_1.
\end{eqnarray}
where again we have used the notation $v_i=\frac{\kappa^2}{4} V_i$.
It is convenient to combine the discontinuity equations to obtain 
the condition for the mechanical balance of the forces at the junction  
\begin{eqnarray}
&& v_2 \sin\varphi_2=v_3\sin\varphi_1, \nonumber \\
&& v_3 \cos\varphi_1+v_2\cos\varphi_2+v_1=0.
\label{eq:balance2}
\end{eqnarray}
These equations are completely determining the angles $\varphi_{1,2}$
by the relation
\begin{eqnarray}
&& \cos\varphi_1=\frac{v_2^2-v_3^2-v_1^2}{2v_1v_3},\nonumber \\
&& \cos\varphi_2=\frac{v_3^2-v_1^2-v_2^2}{2v_1v_2}.
\label{eq:cosine}
\end{eqnarray}
We can now express the 
remaining variables with the help of $m_{1x}$ and $m_{1y}$ 
using the discontinuity equations as
\begin{eqnarray}
& \hspace*{-1.2cm} m_{2x}=m_{1x}-v_1, \; \; \;  & m_{2y}=m_{1y}, 
\nonumber \\ 
& m_{3x}=m_{1x}+v_3\cos\varphi_1, \; \; \;  & m_{3y}=m_{1y}-v_3\sin\varphi.
\end{eqnarray}
Using these expressions the equations in the bulk can be rewritten as
\begin{eqnarray}
&& m_{1x}^2+m_{1y}^2=-\lambda_1, \nonumber \\
&& (m_{1x}-v_1)^2+m_{1y}^2=-\lambda_2, \nonumber \\
&& (m_{1x}+v_3\cos\varphi_1)^2+(m_{1y}-v_3\sin\varphi_1)^2=-\lambda_3.
\end{eqnarray}
From the first two equations $m_{1x}$ can be expressed as
\begin{equation}
m_{1x}=\frac{\lambda_2-\lambda_1+v_1^2}{2v_1}.
\end{equation}
Using this formula, the expression for $m_{1y}$ from the first
equation, and the values of $\cos\varphi$ from (\ref{eq:cosine})
we again obtain a single fine-tuning relation between
the tensions and 
the cosmological constants:
\begin{eqnarray}
&& v_1^2v_2^2v_3^2 +\lambda_1 v_2^2 (v_1^2+v_3^2-v_2^2)
+\lambda_2v_3^2(v_1^2+v_2^2-v_3^2)+\lambda_3v_1^2(v_2^2+v_3^2-v_1^2)+
\nonumber \\
&& v_1^2(\lambda_1-\lambda_3)(\lambda_2-\lambda_3)+v_2^2(\lambda_2-\lambda_1)
(\lambda_3-\lambda_1)
+v_3^2(\lambda_1-\lambda_2)(\lambda_3-\lambda_2)=0.
\end{eqnarray}
In the case of
$\Lambda_1=\Lambda_2=\Lambda_3=\Lambda$ and $V_1=V_2=V_3=V$ this
relation simplifies to $v^2=-3\lambda$, that is   
\begin{eqnarray}
\kappa^2 V^2=-\frac{24}{5} \Lambda.
\end{eqnarray}

\section{Comments on the Cosmological Constant Problem}
\setcounter{equation}{0}
\setcounter{footnote}{0}

One of the biggest puzzles in particle physics 
is the vanishing of the 
cosmological constant (or why its value is
at least 120 orders of magnitudes
smaller than its natural size of the order $M_{Pl}^4$ would
be). There is no symmetry that could forbid the appearance of the
cosmological constant term. Thus the best hope is that there is a
dynamical reason behind the vanishing of the cosmological constant.
However, within four dimensional theories it is very difficult to find 
a dynamical adjustment mechanism that would naturally achieve this
goal (for a review see~\cite{Weinbergreview}).

In the Randall-Sundrum scenario discussed in this paper the vanishing
of the effective four-dimensional cosmological constant is a
consequence of a fine-tuning between the fundamental (5 dimensional)
cosmological constant and the tension of the 3-brane. Thus in the
original RS scenario there is no new information gained about how the
cosmological constant problem could be solved dynamically. 

One can, however, imagine a more complicated scenario like one of the
setups presented in this paper, where the 3-brane we live on arises as
an intersection of different branes. The effective 4 dimensional
cosmological constant is then a function of not only the 5 dimensional
cosmological constant and the brane tensions (including 
the tension of the intersection brane), but also the positions
(angles) of the branes.
Brane configurations considered in this paper (or their most obvious 
generalizations) require at least one fine-tuning in addition to the
adjustment of the angles to set the effective 4 dimensional
cosmological constant to zero. One might hope however, that brane
configurations exist where the effective cosmological constant can be
set to zero by adjusting only the orientations of the branes. In order 
for such a brane-setup to be interesting, the values of the angles of
the  branes
at the point where the effective cosmological constant vanishes also
have to depend on the tension of the 3-brane at the intersection (a
quantity which we did not consider in the models presented in this
paper). This is required so that it is possible to cancel the quantum
corrections to the effective 4 dimensional cosmological constant due
to the fields localized on the intersection by readjusting the
angles of the branes. If such a solution indeed existed, then one could 
translate the cosmological constant problem to a completely dynamical
problem in the given brane setup, that is why the angles of the branes 
are adjusted such that the effective cosmological constant
vanishes. Such a dynamical formulation would be
by itself a useful step
towards the understanding of the cosmological constant
problem. If such a brane configuration indeed existed, one could then
furthermore speculate that the reason for the adjustment of the angles 
to a setup with zero effective cosmological constant is due to the
following mechanism: initially, the positions of the branes are not
adjusted and the effective 4 dimensional cosmological constant
does not
vanish. Therefore, the universe is inflating, thereby exerting
pressure on the branes, which are slowly relaxing towards the static
solution at which the effective 4 dimensional cosmological constant
vanishes. Of course, for this speculative picture to hold, one would
need to investigate the dynamics of the branes (beyond finding a
static brane solution with the described features). 
In this paper we only looked at the particular static
ansatz leading to the 
flat four-dimensional metric. Therefore,
our results only indicate that the point
with the vanishing cosmological constant is the extremum of the potential 
for the angles, but not necessarily the minimum.

From a four-dimensional point of view, the angles of the branes 
appear as scalar fields.
Thus one expects that they need to be light to potentially provide 
a solution to the cosmological constant problem.
Even then one is confronted with the usual problem of the
adjustment mechanisms for solving the cosmological constant problem.
It is difficult to understand why the potential
for one or a few scalars is such
that at the minimum of the potential the cosmological constant
vanishes. Moreover, quantum corrections seem to destroy 
this tuning even if it was true at tree-level. However, it might be
possible, that what seems to be a terrible fine-tuning in the
effective 4 dimensional theory is a simple consequence of brane
dynamics in higher dimensions, with no tuning required in the full
theory of branes 
(after all, if a solution of the desired type existed, 
the value of the cosmological constant in the bulk 
would be generic). If this fine-tuning in the effective theory is
indeed the consequence of brane physics in the higher dimensional
theory, one might hope that it is stable under radiative corrections,
since the quantities that presumably govern the dynamics of the branes 
are the full quantum corrected ones.

In the setup considered here there is another possibility for
improvement on the fine-tuning of the potential in the effective 4
dimensional theory. As we noted,
for a given set of parameters the requirement for the existence 
of the static solutions with the flat four-dimensional metric 
completely determines the angles.
Thus from the four-dimensional point of view, the potential for the 
angles is determined mostly by their interactions with the metric, 
in particular 
with its light KK excitations. The description of the
four-dimensional effective theory in 
the RS configurations includes a large
number of arbitrarily light KK excitations. 
Thus it is not inconceivable that their interactions with the angles
lead to a situation 
qualitatively different from the usual considerations.

\section{Conclusions}
\setcounter{equation}{0}
\setcounter{footnote}{0}

In this paper we have presented static solutions to Einstein's
equations corresponding to branes at angles intersecting in a single
3-brane. Such solutions might be useful for building models with
extra dimensions in the Randall-Sundrum scenario. The 
solutions are obtained by gluing patches of AdS space together, with
the boundaries given by the branes. We find, that a static solution 
of this sort is only possible if the forces from the brane tensions
acting on the junction exactly balance. In addition to this condition 
we find other constraints that the parameters of the theory (the brane 
tensions, angles of the branes and the bulk cosmological constant)
have to satisfy. In all the examples considered in this paper there is 
one fine-tuning relation which is independent of the angles of
the branes and thus can not have a dynamical origin. 
It would be very important to understand, whether or not static brane
configurations of this sort (where all tuning conditions can be satisfied  
by adjusting the positions of the branes) do exist, 
and if so whether they
can be minima of the scalar potential of the angles in the effective 4 
dimensional theory.

\section*{Acknowledgements}
We would like to thank Nima Arkani-Hamed, Tom Banks, and Michael Dine for 
many useful discussions on the cosmological constant problem.
We are also grateful to 
Michael Graesser, Barak Kol,
Chris Kolda, Martin
Schmaltz, Raman Sundrum, and John Terning for helpful discussions and
comments. We thank the Aspen Center for Physics 
(where this work has been initiated) 
and the organizers of the workshop 
``Phenomenology of superparticles and superbranes'' for their
hospitality.  The work of C.C. is 
supported in part by the U.S. Department of Energy under Contract
DE-AC03-76SF00098 and in part by the National Science Foundation under
grant PHY-95-14797. The work of Y.S. is supported in part by NSF
grants PHY-9802484 and PHY94-07194.


\end{document}